\documentclass[preprint]{aastex62}

\usepackage{enumitem,amssymb}
\newlist{todolist}{itemize}{2}
\setlist[todolist]{label=$\square$}
\usepackage{pifont}

\usepackage{soul}
\usepackage{colortbl}
\graphicspath{{./}{figures/}}

\received{January 17th, 2019}
\revised{February 8th, 2019}
\accepted{February 12, 2019}
\submitjournal{ApJL}

\shorttitle{Co-orbital Asteroids as the Source of Venus's Zodiacal Dust Ring}
\shortauthors{Pokorn\'{y} and Kuchner}

\begin{document}

\title{Co-orbital Asteroids as the Source of Venus's Zodiacal Dust Ring}

\correspondingauthor{Petr Pokorn\'{y}}
\email{petr.pokorny@nasa.gov}

\author{Petr Pokorn\'{y}}
\affiliation{Department of Physics, The Catholic University of America, Washington, DC 20064, USA}
\affiliation{Heliophysics Science Division, NASA Goddard Space Flight Center, Greenbelt, MD 20771}
\affiliation{Astrophysics Science Division, NASA Goddard Space Flight Center, Greenbelt, MD 20771}
\author{Marc Kuchner}
\affiliation{Astrophysics Science Division, NASA Goddard Space Flight Center, Greenbelt, MD 20771}

\begin{abstract}

Photometry from the \textit{Helios} and \textit{STEREO} spacecraft revealed regions of enhanced sky surface-brightness suggesting a narrow circumsolar ring of dust associated with Venus's orbit. We model this phenomenon by integrating the orbits of 10,000,000+ dust particles subject to gravitational and non-gravitational forces, considering several different kinds of plausible dust sources. We find that only particles from a hypothetical population of Venus co-orbital asteroids can produce enough signal in a narrow ring to match the observations. Previous works had suggested such objects would be dynamically unstable. However, we re-examined the stability of asteroids in 1:1 resonance with Venus and found that $\sim$8\% should survive for the age of the solar system, enough to supply the observed ring.

\end{abstract}
\section{Introduction}
\label{sec:introduction}
Photometry from the \textit{Helios} spacecraft \citep{Leinert_Moster_2007} and images from the \textit{STEREO} spacecraft \citep{Jones_etal_2013,Jones_etal_2017} have revealed a surface brightness pattern consistent with a circumsolar ring of enhanced dust density at the orbit of Venus. An analogy with the Earth's resonant dust ring \citep{Jackson_Zook_1989,Dermott_etal_1994,Reach_etal_1995} suggests that this pattern could represent sunlight scattered by dust migrating inward from the asteroid belt under radiation drag and detained in Venus's exterior mean motion resonances. Dynamical models of terrestrial-mass planets interacting with circumstellar dust imply that such resonant rings should be common and useful diagnostics of dust and planet properties \citep{Kuchner_Holman_2003,Stark_Kuchner_2008}.

Initial attempts to dynamically model the formation of such a ring of Zodiacal dust created by Venus did not reproduce this ring \citep{Jeong_Ishiguro_2012}. This model used 720 particles, released from a source population representing Jupiter-family comets (JFCs) and main belt asteroids (MBAs). Only a few percent of the particles were trapped in mean motion resonances (MMRs) with Venus, and the model yielded a negligible density enhancement in the vicinity of the ring. Any dynamical model of this phenomenon must overcome Poisson noise in six-dimensional phase space \citep{MoroMartin_Malhotra_2002}, so such small trapping probabilities present a challenge.

To address this situation and search for a plausible explanation for the observed ring, we performed a new set of integrations incorporating 10,000,000+ dust particles responding to radiation pressure, radiation drag, solar wind drag and gravitation from all eight planets. We also considered several different source populations for the dust, summarized in Table 1: JFCs, MBAs, Oort Cloud Comets (OCCs), Halley-type Comets, the young ($< 1$ Ma) asteroid breakup events, and hypothetical families of asteroids in resonant lock with Venus. We collected the positions output by the \texttt{swift\_RMVS\_3} integrator \citep{Levison_Duncan_2013} with radiative forces included \citep{Nesvorny_etal_2011JFC} in a histogram that models the 3-D density distribution of the dust from each source. Each particle was weighted by a factor of $r^{-2}$, where $r$ is the distance to the Sun, to account for solar illumination and a scattering phase function that is smooth on the scale of the observed Venus ring. We used an initial grain size distribution proportional to a power-law, $dN \propto s^{-\alpha} ds$, where $s$ is the grain radius and $\alpha$ is the differential size index. We tested different values of $\alpha \in [2.5,5]$, where $\alpha=3.5$ represents an infinite collisional cascade \citep{Dohnanyi_1969}. Then we integrated these distributions over various lines of sight through the cloud and compared the resulting surface brightness distributions to the \textit{STEREO} and \textit{Helios} observations. 
\begin{deluxetable*}{lccc}
\tablewidth{0pt}

\tablecaption{Summary of source populations, the parent bodies producing the dust in our models.}
\tablenum{1}
\tablehead{\colhead{Source Population} & \colhead{Total number of } & \colhead{Maximum Enhancement at } & \colhead{Reference for population } \\ 
\colhead{} & \colhead{ simulated particles} & \colhead{0.723 au vs 0.7/0.75 au [\%]} & \colhead{description} } 

\startdata
Jupiter Family Comets (JFCs) & 4,487,700 & -0.3\% / 0.4\%  & \citet{Nesvorny_etal_2011JFC} \\
Main-belt Asteroids (MBAs) & 7,182,800 & 7.6\% / -7.9\% & \citet{Nesvorny_etal_2010} \\
Oort Cloud Comets (OCCs) & 160,000 & 2.2\% / 1.0\% & \citet{Nesvorny_etal_2011OCC} \\
Halley-type Comets & 240,000 & 0.9\% / -2.0\% & \citet{Pokorny_etal_2014} \\
Young asteroidal break-up events & 5 $\times$ 445,450 & 7.5\% / -7.3\% & \citet{Nesvorny_etal_2015} \\
Hypothetical Venus co-orbitals & 1,912,300 & 85\% / 117\% & see supplementary  \\
 & & & material\\
Hypothetical Asteroids in 2:3V & 100,000 & 5.9\% / -4.9\% & see supplementary  \\
Mean Motion Resonance & & & material\\
Hypothetical Asteroids in 9:10V & 100,000 & 1.6\% / 5.2\% & see supplementary  \\
Mean Motion Resonance & & & material\\
\enddata

\end{deluxetable*}

\section{Results}
Figure 1A shows the azimuthally-averaged surface brightness profiles for dust released by the first five source populations in Table 1 (i.e. not the hypothetical populations). The profiles are normalized arbitrarily to separate them on the plot. Also plotted are power law fits to each model of the form $r^\delta$, where $r$ is the distance to the Sun. Each of these models is smoothed using boxcar smoothing with a window size of 0.03 au, which is smaller than the scale of the observed Venus ring (0.06 au), and is well approximated by a power law. 

\begin{figure}
\figurenum{1}
\plotone{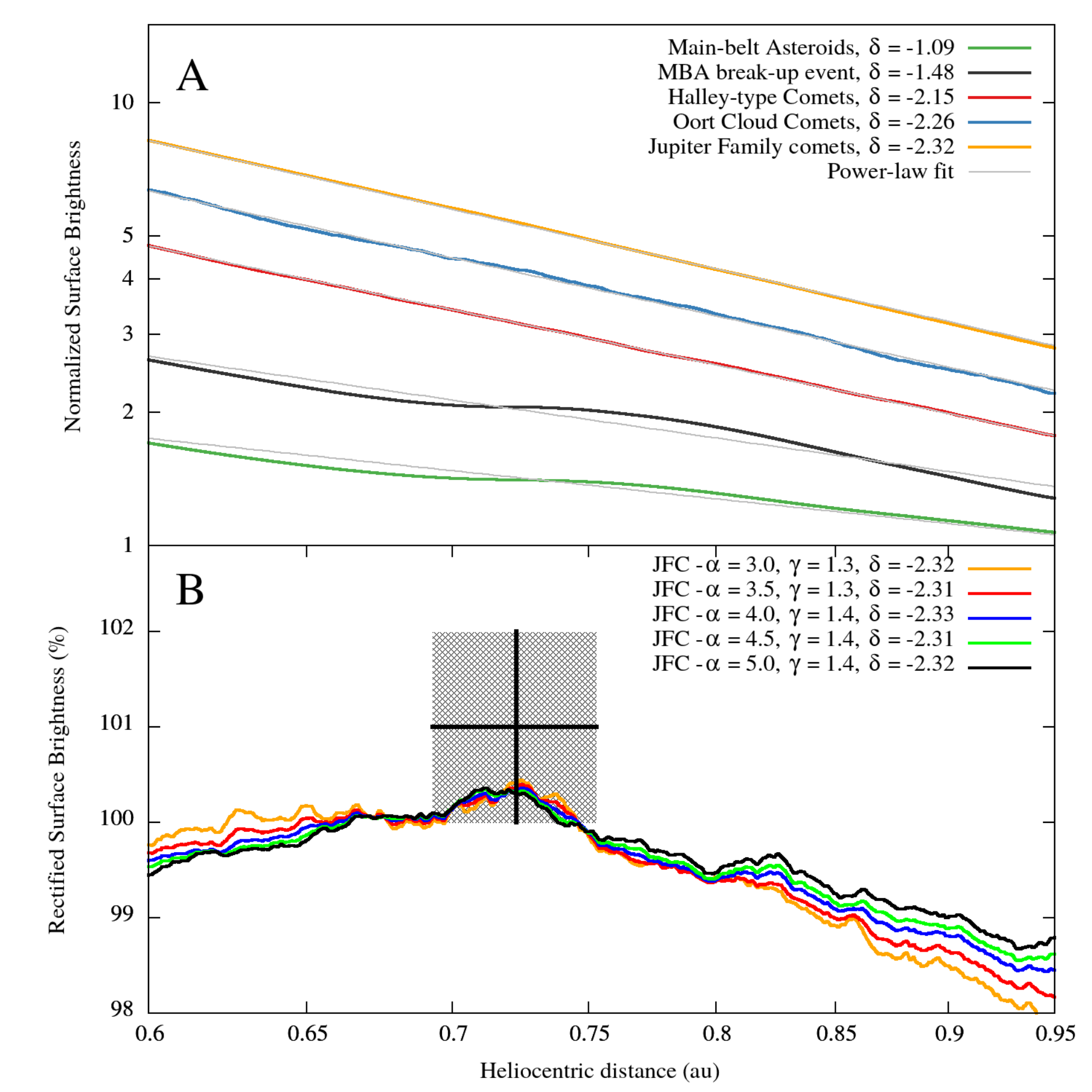}

\caption{Dust surface brightness profiles for different dust producing populations in the inner solar system, normalized arbitrarily for clarity. A) The five major sources of dust in the inner solar system (see Table 1). B) Five different versions of the JFC model divided by the radial power law measured by the \textit{Helios} spacecraft \citep{Leinert_etal_1981} and compared to the Venus ring parameters (grey box with cross) derived from \textit{Helios} photometry by \citep{Leinert_Moster_2007}. Although JFCs dominate the dust population, the surface brightness enhancement they create at the location of the observed ring is roughly 5 times too faint to explain the \textit{Helios/STEREO} observations. }
\end{figure}

\subsection{Jupiter-Family Comets: The Dominant Source of Zodiacal Dust}
JFCs are likely the dominant source of the Zodiacal Cloud dust, supplying $\sim$90-95\% of the grains \citep{Nesvorny_etal_2011JFC}. Supporting this notion, we find that the radial surface brightness distribution associated with dust from JFCs agrees with that measured by the \textit{Helios} spacecraft. Following \citet{Nesvorny_etal_2011JFC}, we weighted the dust production rates of the JFCs by a factor $W=q^{-\gamma}$, where $q$ is the pericenter distance of each comet's orbit. The \textit{Helios} spacecraft measured that the Zodiacal dust surface brightness decreases with the heliocentric distance as $r^\delta$, where $\delta=-2.3\pm0.1$. The JFC source population reproduces this slope given a weighting factor power law in the range $\gamma=1.3\pm0.1$. Figure 1A shows that besides the JFC dust, the OCC grains also yield a steep enough surface brightness slope to plausibly match the power law measured by the \textit{Helios} spacecraft \citep{Leinert_etal_1981}. However, the latitudinal shape of the Zodiacal cloud modeled in \citet{Nesvorny_etal_2011JFC}, combined with the meteoroid ablation observations at Earth \citep{CampbellBrown_2008,Janches_etal_2015,CarrilloSanchez_etal_2016} rule out OCC particles as a major component of the Zodiacal cloud. 

Although JFCs are the dominant source of Zodiacal dust, they do not appear to be the primary source of the Venus ring. Figure 1B compares five different versions of JFC dust model with various weighting factors $(\gamma)$ and size distributions $(\alpha)$ with the parameters of the ring inferred from observations. \textit{Helios} photometry \citep{Leinert_Moster_2007} indicated a surface brightness enhancement at the orbit of Venus with 2\% amplitude and a full width at half maximum (FWHM) of 0.06 au, depicted by the crosshairs and checkered box. These ring parameters are consistent with those derived from \textit{STEREO} observations by \citet{Jones_etal_2017}. The JFC models have been divided by the \citet{Leinert_Moster_2007} power law ($\delta=-2.3$), revealing Poisson noise plus small peaks near Venus, roughly five times too weak to match the height of the checkered box. JFC particles probably cannot themselves reproduce the 2\% enhancement at heliocentric distances close to Venus's orbit, a result that appears to be independent of the size distribution and dust production weighting factor. 

\subsection{Other Plausible Sources for the Dust in the Ring}
Since the JFCs appear unlikely to be the dominant source of the Venus ring, we examined several other potential source populations. In addition to the JFCs and OCCs mentioned above, Figure 1A shows radial surface brightness profiles for dust released from Halley-type comets, MBAs, and a model for young asteroidal break-up events. Table 1 provides references for the distributions we assumed for each of these real small-body populations. Each of these additional small body populations also yields a dust cloud that is well modeled by a radial power law; these power laws are depicted in Figure 1A. In addition to these known sources of dust, we also examined some hypothetical sources of dust: hypothetical populations of asteroids in mean motion resonances (MMRs) with Venus. Few asteroid searches have examined these regions because of the challenge of pointing telescopes toward the sun. We constructed these populations by placing bodies at the nominal semimajor axes of these MMRs and integrating their orbits subject to gravitational perturbations from all eight planets to check them for stability. Since these hypothetical source populations are located at or just beyond Venus's orbit, one might expect them to yield substantial dust density enhancements at Venus's orbit.

Figure 2A compares models of dust from seven real and hypothetical source populations (not including JFCs) to the parameters of the Venus ring as observed by \textit{Helios}. A checkered box like the one in Figure 1B depicts the ring parameters inferred by \citet{Leinert_Moster_2007}. In order to explain the observations of the Venus ring with dust from one of these lesser source populations, it would need to create at least a 20\% peak at the orbit of Venus, to be diluted by the dominant, featureless background of dust from JFCs, which we will conservatively assume contains 90\% of the surface brightness at the orbit of Venus. To illustrate the effect of this dilution, we divided each model by a radial power law of $r^{-2.3}$ and then normalized each model to yield a 10\% contribution just outside the checkered box. 

\begin{figure}
\figurenum{2}
\plotone{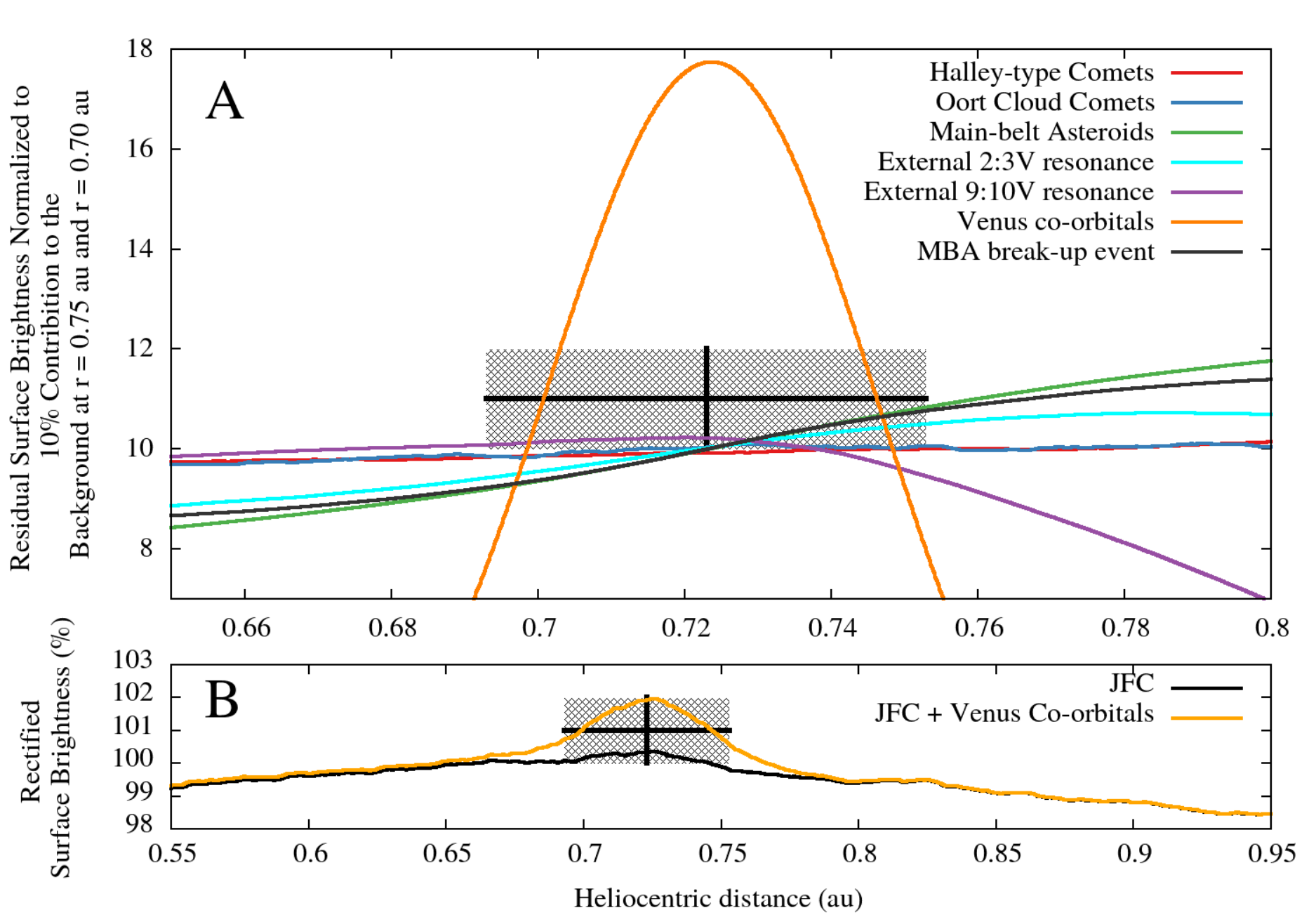}

\caption{Surface brightness distributions of dust from various source populations (curves) compared to the Venus ring parameters derived from \textit{Helios/STEREO} data, illustrated by a checkered box at $r = 0.723$ au with width 0.06 au. All surface brightnesses were divided by a radial power law of $r^{-2.3}$. A) Seven different real and hypothetical source populations normalized to yield a 10\% contribution just outside the checkered box. Although all three models of dust released from bodies in resonance with Venus show some radial structure at 0.72 au, only the model of dust arising from Venus co-orbitals yields a strong enough peak to match the \textit{Helios} ring parameters. B) Our preferred model for the dust environment of Venus, combining dust released from JFCs and Venus co-orbitals in proportions scaled to match the \textit{Helios/STEREO} observations. }
\end{figure}

Figure 2A shows that our models for dust from MBAs, Halley-type Comets, OCCs, and young main belt asteroid breakup events all fail to produce a clear peak at 0.723 au as demanded by the observations. We also examined hypothetical populations of asteroids in all first order exterior MMRs with Venus from 1:2V to 9:10V and also 11:12V, 14:15V and 19:20V. The dust from the hypothetical asteroid population in 2:3V MMR shows a slight decrease in this power-law normalized surface brightness interior of Venus's orbit, and the dust from the hypothetical asteroid population in 9:10V MMR shows a downward trend beyond this radius. However, none of the first order MMRs we examined yields a clear peak at 0.723 au (see supplementary material). 

One model stands out in Figure 2A as producing a strong peak with parameters matching those of the observed ring: the model of dust released by a hypothetical population of Venus co-orbitals. Such co-orbital asteroids would orbit in 1:1 resonance with Venus, at a semimajor axis roughly equal to that of Venus. The third column in Table 1 provides a more quantitative comparison between the models and the data. It lists the maximum surface brightness enhancement each model yields at 0.723 au as compared to both 0.7 au and 0.75 au. We calculate these enhancements by first averaging the surface brightness of each model over heliocentric distances of $0.7\pm0.001$ au, $0.723\pm0.001$ au, and $0.75\pm0.001$ au to mitigate the Poisson noise that arises in the models at small scales; the residual Poisson noise is negligible for our purposes. These enhancements are maximum enhancements in the sense that they conservatively assume that each of these source populations contributes 100\% of the cloud at the orbit of Venus (we expect that they contribute no more than about 10\%). The maximum enhancements in Table 1 show that the only source population we explored that produces a sufficient peak at Venus's orbit is the co-orbitals. Although dust from MBAs, asteroid break-up events and asteroids in the 2:3V MMR yield enhancements (7.5 \% and 5.9 \%) at the center of the ring with respect to the surface brightness at $r = 0.7$ au, each of these dust cloud models contains a decrement in the range of -7.3 \% and -4.9\% on the other side of the observed ring. Dust from asteroids in the 9:10V MMR also creates a lopsided pattern. In contrast, the dust from the co-orbitals yields a strong, roughly symmetrical peak centered right at Venus's orbit.

The reason only Venus co-orbitals are so effective at producing a ring structure is that the normal process of resonant trapping of Zodiacal dust in first-order MMRs \citep[e.g.,][]{Jackson_Zook_1989,Dermott_etal_1994,Stark_Kuchner_2008} is highly inefficient for 1:1 resonances \citep{Liou_Zook_1995}. This inefficiency stems from the approximate conservation of the Jacobi constant in the vicinity of Venus. Dust particles larger than $D \simeq 60 ~\mu$m ejected from Venus co-orbitals are born with a low value of the Jacobi constant and remain in the 1:1 MMR, producing the co-orbital ring illustrated here. Dust particles smaller than 60 μm are kicked by radiation pressure outside the 1:1 MMR, and their dynamical evolution follows the same pattern as the other populations migrating from the outer parts of Venus's orbit. In contrast, particles migrating via Poynting Robertson drag from substantially outside of 1:1 resonance are created with higher values of the Jacobi constant, and are not permitted to have orbits with guiding centers near Venus's Lagrange points. Consequently, as such particles are dragged past Venus, they must pass near Venus, where they are either ejected to more distant orbits or scattered into orbits interior to Venus's Hill sphere; this process produces the decrements seen in the models of dust from MBAs, young asteroid break-up events and asteroids in the 2:3V MMR. 

\subsection{Our Preferred Model: Dust from Venus's Co-Orbitals}
Based on the patterns described above, our preferred model for the dust environment of Venus, combines dust released from JFCs and Venus co-orbitals. Figure 2B illustrates this preferred model as viewed by \textit{Helios}. To match the 2\% increase of surface brightness inferred from \textit{Helios} observations, as shown in Figure 2B, this model predicts a 9.5\% maximum overdensity of dust in the ring, consistent with the $\sim$8.5\% maximum overdensity estimated from \textit{STEREO} observations \citep{Jones_etal_2017}.

\begin{figure}
\figurenum{3}
\epsscale{0.9}
\plotone{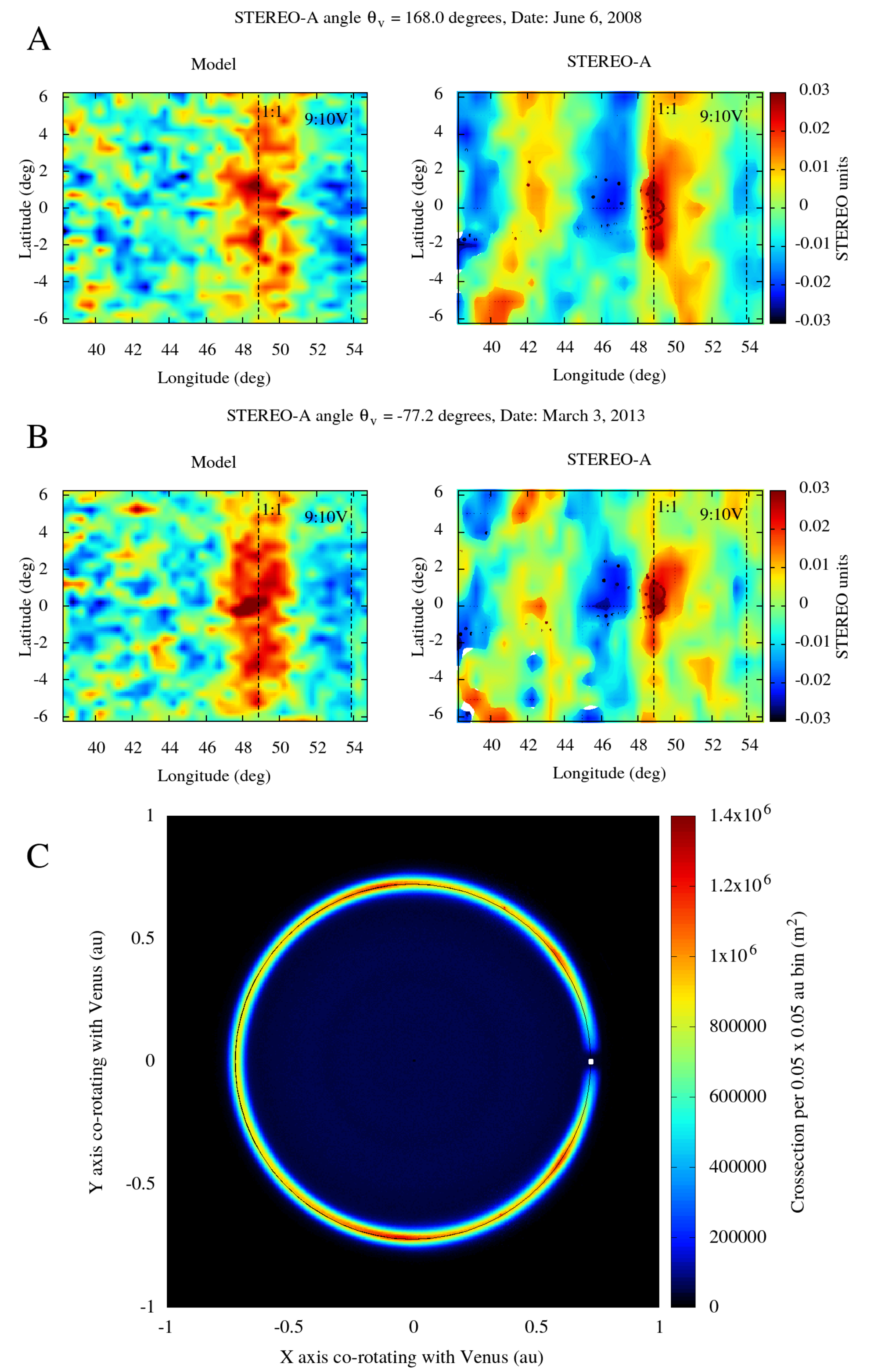}

\caption{Model of the co-orbital dust compared to \textit{STEREO} images (A and B) and viewed face-on (C). Vertical lines in A and B indicate the locations of the 1:1 and 9:10V mean motion resonances for large dust grains. A feature associated with the 2:3V resonance would lie at a longitude of roughly 81 degrees, off to the right of these images. The face-on view (C) shows that the co-orbital dust model contains azimuthal structure that might explain the azimuthal structures in the Venus ring observed by \citet{Jones_etal_2017}.}
\end{figure}

Figure 3A and 3B illustrates this preferred model as viewed by \textit{STEREO} compared to corresponding views of the Venus ring as seen by \textit{STEREO}, from \citet{Jones_etal_2017}. Our model does not capture the tilt of Venus's orbit because it averages over the precession period to beat down the Poisson noise, so the model lies in the ecliptic plane, while the signal seen in the \textit{STEREO} data is centered at the latitude of Venus at that epoch. However, our model does agree with the surface brightness enhancements seen by \textit{STEREO} in radial width (0.06 au) and in height above the ecliptic (0.1 au). We removed a smoothly varying background from these images following the same procedure \citet{Jones_etal_2013,Jones_etal_2017} followed to highlight the ring in the \textit{STEREO} data. Note that our simulation also reproduces the over-subtractions (blue regions) appearing to the left and to the right of the ring in the \textit{STEREO} images. We compare our model to observations that are 5 years apart and probe different sections of the ring and find reasonable agreement. 

Figure 3C shows our preferred ring model as viewed from above the ecliptic plane. The wealth of imaging data from the two \textit{STEREO} spacecraft suggests that the ring has significant azimuthal structure \citep{Jones_etal_2017}. Likewise, the face-on view of our model (Figure 3C) shows four distinct azimuthal peaks in surface density: one pair associated with the L4 Lagrange point, one pair associated with the L5 Lagrange point. Each pair represents the extremes of the libration cycle for the corresponding 1:1 resonance. The details of this azimuthal structure likely depend on the details of the co-orbital source population, which is presently unknown. Another feature evident from our model is the slight prograde shift of the L4 and L5 locations caused by radiation drag, described analytically by \citet{Murray_1994}.

To obtain the total mass of the Venus dust ring according to our model, we use the total cross section of the Zodiacal Cloud beyond 1 au, $\Sigma_\mathrm{ZC}=2 \times10^{11}$ km$^2$ \citep{Nesvorny_etal_2011JFC} as a calibration point. Similar to \citet{Nesvorny_etal_2011JFC} we assume that the bulk of the Zodiacal Cloud is comprised of JFC particles. Then we calculate two values from our JFC model: the total particle cross-section beyond 1 au, and the total particle cross-section between 0.7 and 0.75 au. By scaling our model to match $\Sigma_\mathrm{ZC}$, and assuming a Gaussian enhancement in surface brightness with an amplitude of 2\% we obtain the total particle cross-section of the ring structure, $\Sigma_\mathrm{ring}=4.3 \times 10^7$ km$^2$. We then obtain the mass of all particles in the dust ring provided from our model using a standard conversion between the total cross-section and total mass \citep[see e.g.,][]{Pokorny_etal_2014}. For a bulk density of 3 g cm$^{-3}$ we predict the total mass of the ring to be $1.3_{-0.6}^{+2.9} \times 10^{13}$ kg, which translates to an asteroid with radius $R=1015_{-198}^{+490}$ m, i.e., a single 2 km asteroid ground into dust.

\added{All dust particles assumed in our model and captured in the 1:1 MMR have Poynting-Robertson drag timescales of less 1 Myr. This means that the observed co-orbital dust ring is either recently created, or continuously replenished, like the rest of the zodiacal dust.}
The dust could arise from the slow grinding of a collisionally evolved population of asteroids, or it might indicate a recent asteroidal collision, akin to the Karin and Veritas asteroid families and their corresponding dust bands \citep{Nesvorny_etal_2006}. For comparison, the dust associated with the Karin family has total inferred cross section in the range of $1300-2500 \times10^7$ km$^2$ , corresponding to a single 20-40 km asteroid ground into dust and the dust associated with the Veritas family has total inferred cross section in the range of $2500-5000 \times10^7$ km$^2$ , corresponding to a single 20-50 km asteroid ground into dust; the ranges represent different assumptions about the size frequency distribution of the dust.  \deleted{Either scenario points to a larger population of co-orbital asteroids. }

\section{Discussion: Stability of Venus Co-orbitals}
The stability of Venus's co-orbital asteroids (Trojans and horseshoe librators) has been a matter of some debate. Early investigations suggested that all of Venus's co-orbital asteroids would be unstable \citep{Scholl_etal_2005}. Yet a recent work integrated the orbits of Venus Trojans for 700 Myr \citep{Cuk_etal_2012}, and found that a substantial fraction were stable over this time period.

To check that the source population favored by our dust models would indeed be stable over the entire lifetime of the solar system, we performed our own orbit integrations, using \texttt{swift\_RMVS\_4} code \citep{Levison_Duncan_1994}, and gravitational perturbations from all 8 planets. We constructed an initial population of particles in co-orbital resonance using short integrations, assembling a group of 10,000 particles that were stable near Venus for 10 Myr.  We then integrated the orbits of these particles for the lifetime of the solar system (4.5 billion years).  Figures 4A and 4B show the results of these simulations; it displays the initial orbital elements of the particles we tested, with various symbols indicating the times when the particles were ejected from the simulation when their semimajor axes became $a < 0.7088$ au or $a > 0.7378$ au, i.e., more than 2\% different from the semimajor axis of Venus. Of the 10,000 particles, 8.2\% remained stable for the whole simulation of 4.5 billion years, confirming the notion that indeed primordial Venus co-orbitals could still exist today. 
\begin{figure}
\figurenum{4}
\plotone{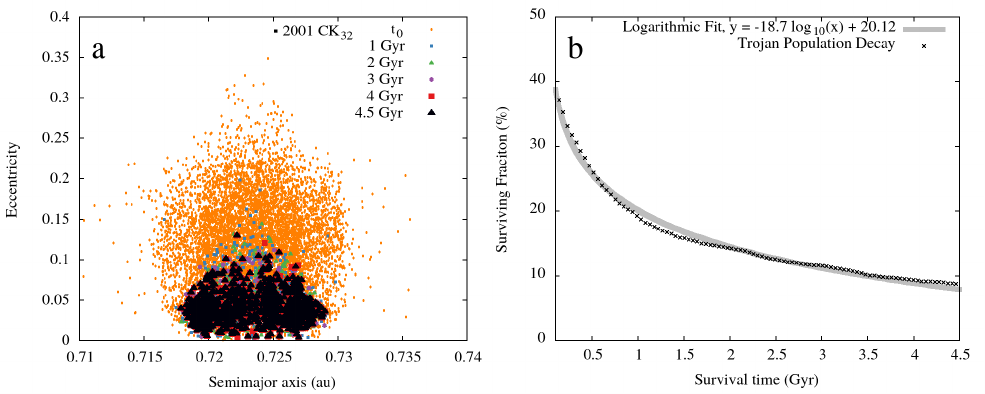}

\caption{Simulations of the stability of Venus co-orbitals. (A) The initial semimajor axis vs. the initial eccentricity for simulated asteroids. The points are color-coded by the survival time. The known Venus co-orbital with the lowest orbital eccentricity (2001 CK32) appears at the very top of this plot, with semimajor axis, $a= 0.725266$, au and eccentricity $e= 0.382614$. (B) The percentage of surviving co-orbitals vs time in our integrations, well modeled by a logarithmic decay. Eight percent of these low-eccentricity objects survive for the lifetime of the solar system. }
\end{figure}

There are currently four or five known Venus co-orbitals: 2001 CK32, 2002 VE68, 2012 XE133, 2013 ND15, and possibly 2015 WZ12 \citep{delaFuenteMarcos_delaFuenteMarcos_2017}. However, these objects are not representative of the population we have in mind to explain the Venus ring. These known co-orbitals are likely known because their large eccentricities, ranging from 0.383 to 0.612, make them near-Earth asteroids, and hence relatively easy to observe without concern for small solar angles. At the same time, their high eccentricities make them unlikely to be stable in their current orbits for the lifetime of the solar system (see Figure 4B). We propose that an undetected population of stable, primordial low-eccentricity ($e < 0.1$) Venus co-orbitals is responsible for the Venus dust ring. 

\added{The possible existence of a substantial population of Venus co-orbitals raises the question of how this population compares to the population of Earth co-orbitals. Observation of Earth co-orbitals has been quite challenging as described in \citet{Wiegert_etal_2000} and \citet{Morais_Morbidelli_2002} due to the geometry of the observation and the existence of the atmosphere at Earth.  However, {\it Gaia} is expected to detect any Earth Trojans with diameters larger than 600 m \citep{Todd_etal_2014}.

The amount of damping in terrestrial planet formation remains unknown, \citet{Morais_Morbidelli_2002} argued that ``\emph{The violent nature of the currently believed scenario for the formation of the terrestrial planets \citep{Chambers_Wetherill_1998} is not favorable to the existence of a significant primordial population of Trojans asteroids associated with them.}''  However, since then, models of planet formation incorporating gas drag, and the discovery of abundant exoplanets in (near) resonant orbits point to a gentler formation process \citep[e.g.][]{Lee_etal_2013,Ramos_etal_2017}, which could be friendlier to primordial co-orbitals.  Comparing the populations of Venus's, Mars's and Earth's co-orbitals should yield constraints on the amount of dissipation during the formation process.}

\acknowledgments
\noindent Funding: P.P. and M.J.K. were supported by NASA Solar System Workings grant NNH14ZDA001N-SSW. \\
Author contributions: 
P.P.: Conceptualization, formal analysis, investigation, methodology, software, validation, visualization, writing - original draft, writing - review \& editing \\
M.J.K.: Conceptualization, formal analysis, funding acquisition, investigation, methodology, validation, writing - original draft, writing - review \& editing\\
Competing interests: Authors declare no competing interests.\\
Data and materials availability: All data is available in the main text or the supplementary materials.

\appendix
\title{Supplementary material for: Co-orbital Asteroids as the Source of Venus's Zodiacal Dust Ring}
\renewcommand{\thefigure}{A\arabic{figure}}

\section{Model overview}
Here we briefly describe all dust population models used in this manuscript. All particles in our models are released from their parent objects described below. Once released from their parent bodies particle dynamics are influenced by gravitation of all eight planets and the effects of radiation pressure, Poynting-Robertson drag and solar wind \citep{Burns_etal_1979}. The solar wind is treated as a correction factor of 1.3 multiplying the Poynting-Robertson drag force \citep{Gustafson_1994}. The effects of solar radiation pressure can be quantified using a single dimensionless parameter $\beta = 11.5 \times 10^{-5}  Q_\mathrm{pr} / (\rho D)$, where $Q_\mathrm{pr}$ is the radiation pressure coefficient, $\rho$ is the particle bulk density and $D$ is the particle diameter, where all units are in cgs units. We use $Q_\mathrm{pr}=1$, which corresponds to the geometric optics limit for particles much larger than the incident light wavelength. When the grains are created, their initial orbits conserve their birth velocities, while suddenly feeling a central force decreased by a factor of $(1 - \beta)$ because of radiation pressure. This effect tends to place them on initial orbits with higher eccentricities and larger semimajor axes than their parent bodies. The initial orbital parameters for all modeled particles are available upon request.

\subsection{Jupiter Family Comets}
The initial orbital distribution of Jupiter Family Comets is taken from \citet{Nesvorny_etal_2011JFC} who employed results from \citet{Levison_Duncan_1997} who followed the orbital evolution of Kuiper belt objects scattered by interaction with planets. A small fraction of these scattered Kuiper belt objects evolved into the inner solar system. Once such an object reaches a perihelion distance $q$ that is smaller than $q_\mathrm{lim}$ we consider it to be part of a pool of dust generating parent bodies. We explored 10 values of $q_\mathrm{lim}$ uniformly distributed between 0.25 au and 2.50 au (step 0.25 au), each containing between 200-4500 parent bodies.

For each of these groups of parent bodies, we generated $N = 2,000,000/D$ dust particles in each of six bins, each with a different dust grain diameter $D$, where $D=10.00,14.68,31.62,68.13,146.8,316.2 ~\mu$m, for a total of 4,487,700 initial orbits. For the initial conditions of these dust grains, we used a random number generator to pick a combination of initial semimajor axis $a$, eccentricity $e$ and inclination $I$ from the pool of parent bodies, and choose the initial argument of pericenter $\omega$, longitude of the ascending node $\Omega$ and mean anomaly $M$ from a uniform distribution between 0 and 2$\pi$. We assumed a bulk density of $\rho = 2$~g~cm$^{-3}$ for all Jupiter Family comet particles and used an integration time step of 3.65 days.

\subsection{Main Belt Asteroids}
Inspired by \citep{Nesvorny_etal_2010}, we selected the parent bodies for dust particles in this model from among 5820 asteroids larger with diameter $> 15$ km in the ASTORB catalog \citep{Bowell_etal_1994}. In each of eleven grain diameter bins, $D= 10.00, 14.68, 21.54, 31.62, 46.42, 68.13, 100.0, 146.8, 215.4, $ $316.2, 464.2 ~\mu$m, we generated $N = 2,000,000/D$ particles. The initial orbital elements $a,e,I$ for each particle were selected from a random member of the parent body population, whereas the initial argument of pericenter $\omega$, longitude of the ascending node $\Omega$ and mean anomaly $M$ were generated randomly from a uniform distribution between 0 and $2\pi$. We assumed a bulk density of $\rho = 3$~g~cm$^{-3}$ and used an integration time step of 3.65 days.

\subsection{Oort Cloud Comets}
We began by generating initial orbits using following conditions: perihelion distance $q$ chosen at random from a uniform distribution of $0.5 < q < 2.5$ au, initial argument of pericenter $\omega$, longitude of the ascending node $\Omega$ and mean anomaly $M$ chosen at random from a uniform distribution between 0 and $2\pi$ \citep{Nesvorny_etal_2011OCC}. We selected the cosine of the inclination for each orbit from a uniform distribution $-1 < \cos(I) < 1$ in order to obtain and isotropic distribution of $I$. We generated $N = 20,000$ particles for each diameter bin $D=10,20,50,100,200,400,800,1200 ~\mu$m for three different initial semimajor axes: $a=300,1000,3000$ au. This selection then yields the eccentricity $e=1-q/a$, Ultimately we used only $a=300$ au models, since they provided much better statistics than models with larger $a$, and their orbital evolution did not appear to show any significant differences. We assumed a bulk density $\rho = 2$~g~cm$^{-3}$ and used an integration time step of 1 day.

\subsection{Halley-type Comets}
Here we used the particle distribution model from \citet{Pokorny_etal_2014}, who adopted dynamical evolution of scattered disk objects from \citet{Levison_etal_2006}. The semimajor axis, eccentricity, perihelion distance and inclination distribution can be found in Figure 4 in \citet{Pokorny_etal_2014}. The model we used here contains $N=20,000$ particles generated in each of 12 diameter bins, $D=10, 20, 50, 100, 200, 400, 600, 800, 1000, 1200, 1500, 2000 ~\mu$m with bulk density $\rho = 2$~g~cm$^{-3}$. The integration time step was 1 day.

\subsection{Young asteroidal break-up events}
Here we considered parent bodies from five different recent asteroidal break-up events with age $<$ 1 Ma adopted from \citet{Nesvorny_etal_2015}; (1270) Datura, (2384) Schulhof, (14627) Emilkowalski, (16598) 1992 YC2, (21509) Lucascavin. For each of these parent bodies we took their Keplerian elements from AstDys database and simulated $N=1,000,000/D$ particles in each of the following diameter bins, $D= 4.642, 10.00, 21.54, 46.42, 68.13, 82.54, 100.0, 121.1, 177.8, 215.4, 261.0, 316.2 ~\mu$m, assuming the bulk density to be $\rho = 3$~g~cm$^{-3}$. The initial semimajor axis, eccentricity and inclination of each dust particle before ejection was the same as their parent bodies while the initial argument of pericenter $ \omega$, longitude of the ascending node $\Omega$ and mean anomaly $M$ of each grain before ejection was chose from a uniform distribution between 0 and $2\pi$. The integration time step was 3.65 days.

\subsection{Hypothetical Venus co-orbitals}
\begin{figure}[h]
\centering
\includegraphics[width=0.65\textwidth]{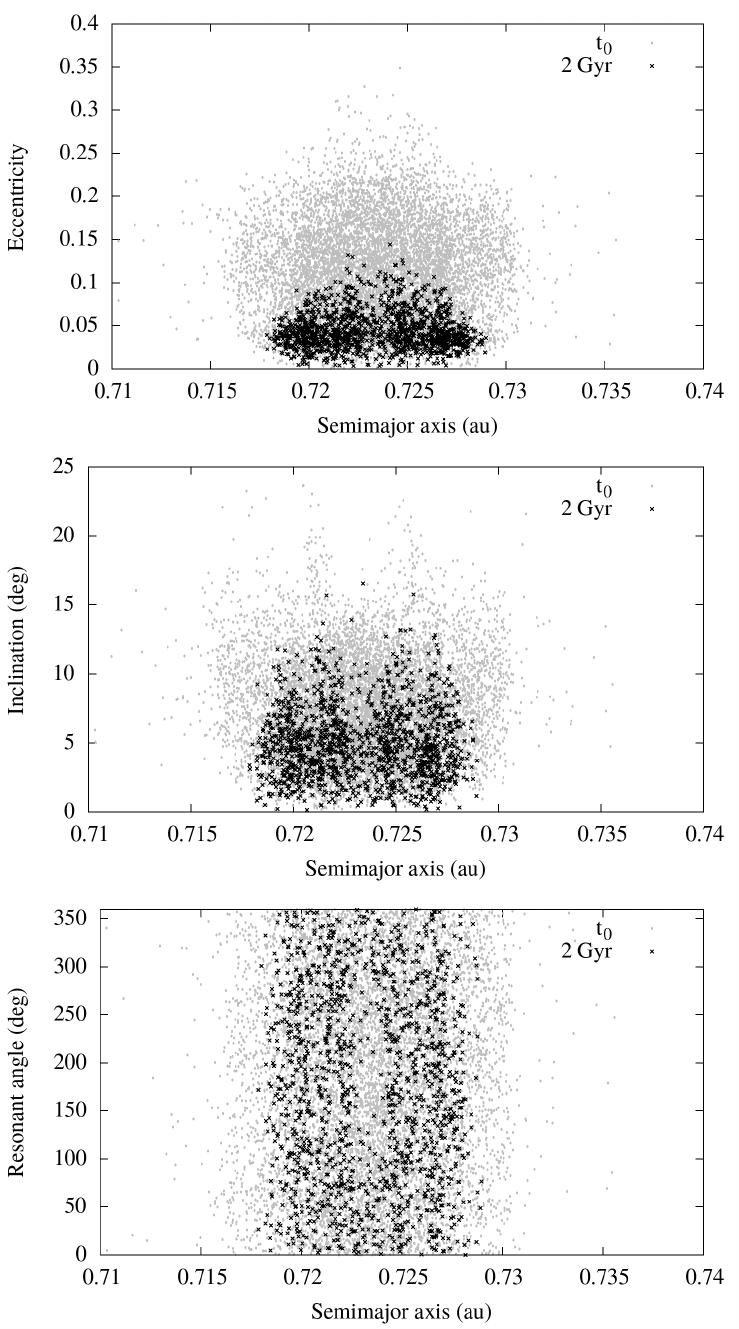}
\caption{Distributions of initial orbits of Venus co-orbitals in our model (gray circles labeled as $t_0$) and the seed population for the dust population of Venus co-orbitals (black crosses labeled as 2 Gyr).}
\label{fig:TROJANS}
 \end{figure}

To examine the stability of Venus co-orbitals and generate a population of co-orbital parent bodies, we began by creating a seed population with initial semimajor axis $a=0.7233$ au, eccentricity in range $0<e<0.2$ and inclination $0^\circ<I<20^\circ$. We integrated the orbits of this seed population for 10 Myr and discarded particles whose semimajor axes failed to remain in the range $0.7088 < a < 0.737$ au. We repeated this process of generating particles, integrating their orbits, and discarding those outside the resonance until we obtained 10,000 particles that survived for 10 Myr.  We considered the 10,000 particles generated in this manner as a representative population of particles in 1:1 resonance with Venus. Figure \ref{fig:TROJANS} shows their orbital elements (after 10 Myr), labeled as $t_0$.

We then began a much longer integration of this group of 10,000 particles, for 4.5 Gyr, to examine what fraction of objects in 1:1 resonance with Venus can survive for the lifetime of the solar system. The results of this longer integration are shown in Figure 4.  

For our models of dust, we selected the initial position and velocity vectors of dust particles randomly from a pool of parent bodies after a period of 2 Gyr. These parent bodies are depicted by black crosses in Figure \ref{fig:TROJANS}.  To ensure that due to limited number of parent bodies particles will not follow exactly the same path we introduced a small, 1~m~s$^{-1}$ velocity kick in a random direction to the initial particle velocity vector. Afterwards, the effects of the radiation pressure were by default included in the numerical integration. We simulated $N=1,000,000/D$ particles in each of the following diameter bins, $D= 1.000, 2.154, 4.642, 10.00, 21.54, 46.42, 68.13, 82.54, 100.0, 121.1, 177.8, 215.4,$ $261.0, 316.2, 383.1 ~\mu$m, and assumed the bulk density to be $\rho = 3$~g~cm$^{-3}$. The integration time step was 3.65 days.

\subsection{Hypothetical Asteroids in 1:2V - 19:20V Mean Motion Resonances}
We examined hypothetical populations of asteroids or large meteoroids in and near all first order exterior mean motion resonances with Venus from 1:2V to 9:10V and also 11:12V, 14:15V and 19:20V. We placed 10,000 parent bodies on orbits with semimajor axes of $a=a_V  [p/(p+1)]^{2/3}$  , where $a_V=0.7233$ au is the semimajor axis of Venus, and $p$ is an integer $\in [1,2,3,4,5,6,7,8,9,11,14,19]$, eccentricities drawn from a uniform distribution between $0<e<0.2$, and inclinations chosen from a uniform distribution between $0^\circ<I<20^\circ$. The initial argument of pericenter $\omega$, longitude of the ascending node $\Omega$ and mean anomaly $M$ of MMR parent bodies were generated randomly from a uniform distribution between 0 and $2\pi$. We integrated the orbits of these parent bodies for 1 Myr with an integration time step of 3.65 days. At the end of the integration we used Eq. (8.76) from \citet{Murray_Dermott_1999} to select parent bodies that remained in MMRs to create the population of parent bodies for our dust model

\begin{figure}[h]
\centering
\includegraphics[width=0.65\textwidth]{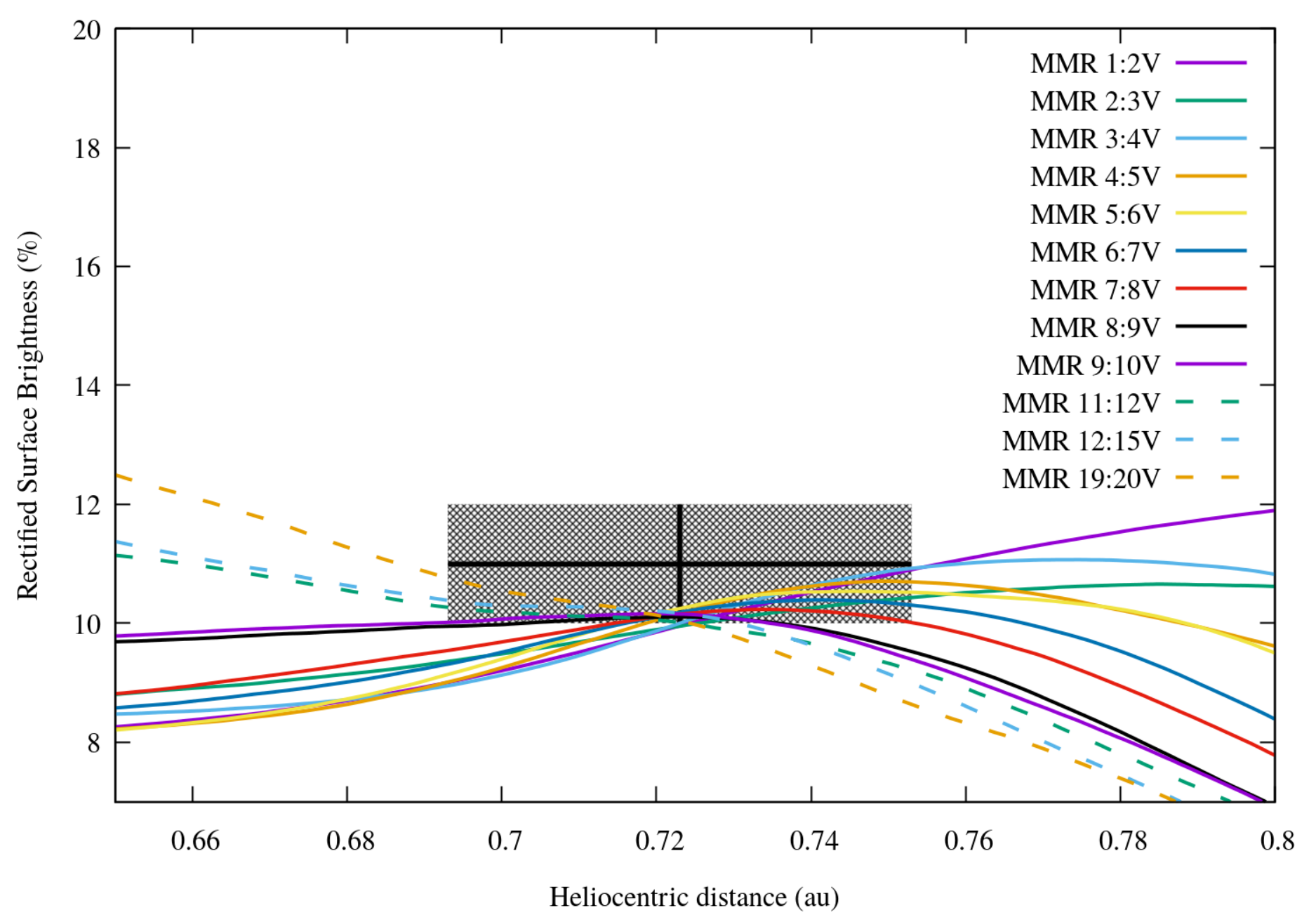}
\caption{Surface brightness distributions of dust from twelve hypothetical resonant source populations (curves) compared to the Venus ring parameters derived from \textit{Helios}/\textit{STEREO} data, which are illustrated by a checkered box at $r = 0.723$ au with width 0.06 au. All surface brightnesses were divided by a radial power law of $r^{-2.3}$.  The dust models originating in exterior MMRs of Venus are all normalized to yield a 10\% contribution just outside the checkered box. }
\label{fig:MMR}
 \end{figure}

For each modeled MMR we released dust particles in each of the following diameter bins: $D= 10.00, 14.68, 21.54, 31.62, 46.42, 68.13, 100.0, 146.8, 215.4 ~\mu$m. We generated $N=100,000/D$ dust grains in each bin, calculating their $\beta$ values assuming a bulk density of $\rho = 3$~g~cm$^{-3}$ and using an integration time step of 3.65 days. Of these, the 2:3V and 9:10V cases are highlighted in the main body of the paper.  However, none of these populations of parent bodies in/near resonances yielded a substantial peak in flux near the ring detected by \textit{Helios} and \textit{STEREO}, which is shown in Figure \ref{fig:MMR}.

\section{Dust Semimajor Axis Distributions}

\begin{figure}[h]
\centering
\includegraphics[width=0.9\textwidth]{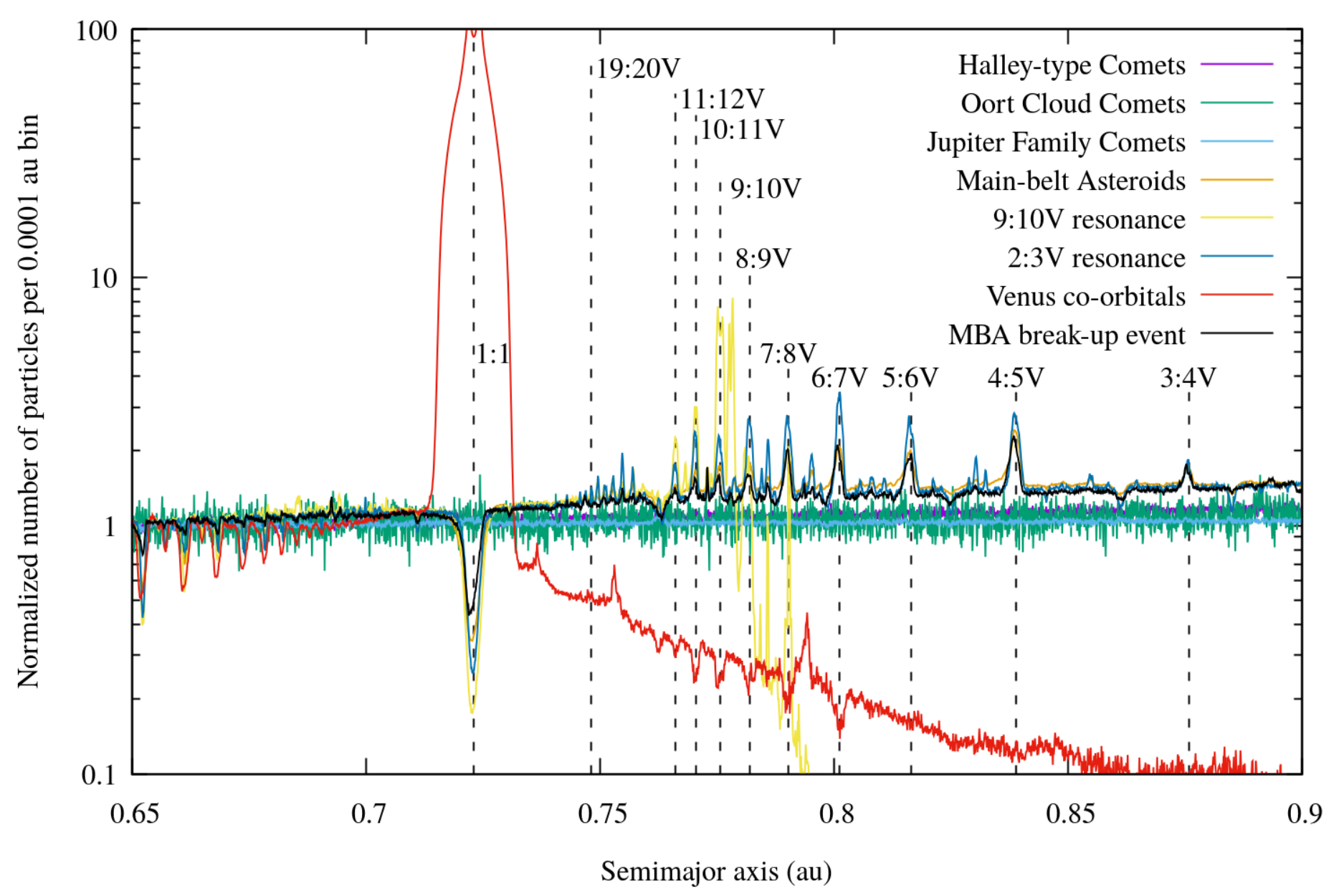}

\caption{Normalized semimajor axis distribution for 8 different dust models, i.e., the number of particles per 0.0001 au bin.  Vertical dashed lines show the locations of several exterior MMRs and the 1:1 MMR. While some models (e.g. the 9:10V MMR, the 2:3V MMR and the MBA breakup event models) show trapping of dust in the exterior MMRs, only the dust generated by co-orbital asteroids yields significant trapping in the 1:1 MMR.}
\label{fig:histA}
 \end{figure}

Figure \ref{fig:histA}. compares the semimajor axis distributions of all of our dust models, weighted according to the size frequency distribution described below.  Mean motion resonances (MMRs), including 1:1 resonances, appear as both peaks and troughs.  Since radiation pressure shifts the locations of the MMRs slightly, some resonances are smeared out; vertical lines indicate the locations of MMRs in the absence of this effect. Nonetheless, this view of the models contains some important features of the gravitational interactions of the dust with Venus. The co-orbital model yields by far the biggest peak, split at the top by libration.  Dust from other sources sometimes becomes trapped in exterior MMRs, but tends to avoid the 1:1 region, as evidenced by the troughs at 0.723 au, particularly in dust from main belt asteroids.

\subsection{Face-on view with respect to the reference frame of a selected planet}
To produce the face-on view shown in Figure 3C, we transformed all the dust grain co-ordinates into a frame rotating with the mean motion of Venus. For Venus, (or any other selected planet) we perform the following conversion to place the planet at  $x_\mathrm{fix}=r_\mathrm{pl}$, and $y_\mathrm{fix}=0$, where $r_\mathrm{pl}$ is the heliocentric distance of the planet:
\begin{eqnarray}
x_\mathrm{fix} &=& x \sin(\varphi) - y \cos{ (\varphi)},\\
y_\mathrm{fix} &=& x  \cos(\varphi) + y \sin (\varphi),
\end{eqnarray}
where $\varphi = \mathrm{atan2}(y/x)+\pi/2$ is the rotation angle and $\mathrm{atan2}$ is the inverse tangent expanded to the $(0,2\pi)$ range. This rotation is then applied at each recorded integration time to the coordinates of all meteoroids, rotating the $(x,y)$ plane so that the selected planet lies on the $x$-axis and all meteoroids are viewed in that planet's co-rotating frame. We do not apply any rotation/offset in the $z$-axis, thus we effectively time-average over the selected planet's oscillations above/below the ecliptic. Furthermore, if the heliocentric distance projection to the ecliptic of the planet changes we do not apply any corrections. This can cause a small radial smearing of disk-like features for a planet on an eccentric orbit.

All meteoroids are binned into $0.075\times 0.075$ au bins thus in our face-on view the value in each bin represents the column density $\Sigma$ of meteoroids in a particular bin. In order to convert the column density Σ to the optical depth τ we use the following expression
\begin{equation}
\tau = \Sigma A \sigma = \Sigma A \pi \overline{r}^2,
\end{equation}
where $A$ is the bin surface area and $\overline{r}$ is the average radius of meteoroids in a particular bin. This equation is only valid if the absorbers do not shadow each other, which is a reasonable assumption for the Zodiacal cloud density in the solar system.
\\
\\
\subsection{Mapping of the flux emitted by meteoroids}

For the scattered light emission of a given meteoroid $\mathcal{E}$ we use a following expression:
\begin{equation}
\mathcal{E}=\pi D^2/4\frac{1}{r_\mathrm{hel}^2} \frac{1}{r_\mathrm{dist}^2} \mathcal{P}(\cos{\gamma}),
\end{equation}
where $r_\mathrm{hel}$ is the heliocentric distance and $r_\mathrm{dist}$ is the distance from the particle, $D$ is the diameter of a meteoroid, $\mathcal{P}$ is the scattering phase function depending on $\cos{\gamma}$, where $\gamma$ is the scattering angle with $\cos{\gamma} = (r_\mathrm{hel}^2 + r_\mathrm{dist}^2 -r_\mathrm{STEREO}^2)/(2 r_\mathrm{dist} r_\mathrm{hel})$, and $r_\mathrm{STEREO}$ being the heliocentric distance of \textit{STEREO} spacecraft. We use a softening parameter $\epsilon=0.01$ for $r_\mathrm{dist}= \sqrt{r_\mathrm{num}^2+\epsilon^2}$ in order to prevent overestimation of individual meteoroids in $\mathcal{E}$. $\mathcal{P}$ is based on {\it \textit{Helios}} observations \citep{Hong_1985}. 

For each position of the spacecraft in our simulation we calculate $\mathcal{E}$ for each meteoroid in our dynamical model. All meteoroids are binned into a 2D map with $0.5^\circ \times 0.5^\circ$ bins in the sun-centered ecliptic longitude ($\lambda$; x-axis) and the ecliptic latitude ($\beta$; y-axis). The reference frame is the same as used in \citet{Jones_etal_2013,Jones_etal_2017}, where the sun is at the origin, and $\beta = 0$ denotes the ecliptic plane.

\subsection{Estimating the cross-section of the Zodiacal cloud beyond 1 au}
In order to calibrate our model we use the cross-section $\Sigma_\mathrm{ZC}$ of the Zodiacal cloud (ZC) as a reference for our calibration. In order to calculate $\Sigma_\mathrm{ZC}$  beyond heliocentric distance $r_\mathrm{hel}$ for each modeled meteoroid we estimate the fraction of time $F_{>r_\mathrm{hel}}$ that meteoroid spends beyond $r_\mathrm{hel}$:

\begin{equation}
F_{>r_\mathrm{hel}} = \frac{\Delta t}{T_\mathrm{orb}} = \frac{\pi - E_* + e\sin E_*}{\pi},
\end{equation}
where $\Delta t$ is the time that the meteoroid spends beyond $r_\mathrm{hel}$, and $T_\mathrm{orb}=2\pi/n$ is the orbital period of the meteoroid with $n$ being the mean motion. In order to calculate  $\Delta t$ we start with the expression of the heliocentric distance $r_\mathrm{hel}$ with respect to the eccentric anomaly $E$:

\begin{equation}
r_\mathrm{hel} = a(1-e \cos E),
\end{equation}
where $a$ is the meteoroid's semimajor axis, and $e$ is the eccentricity. Then the meteoroid's eccentric anomaly at $r_\mathrm{hel}$ is:
\begin{equation}
E_*=\mathrm{arccos}\left( \frac{a-r_\mathrm{hel}}{ae} \right).
\end{equation}
To obtain $\Delta t$ we differentiate Kepler's equation

\begin{equation}
nt= E- e\sin E \Rightarrow n dt = dE (1 -e\cos E)
\end{equation}
and integrate both sides of the differential equation:
\begin{equation}
\frac{1}{2} \Delta t = \frac{1}{2}\int_0^{\Delta t} = \int_{E_*}^{\pi} \frac{1-e\cos E}{n} = \frac{\pi - E_* + e\sin E_*}{n}
\end{equation}
There are two limiting cases for $F_{>r_\mathrm{hel}}$: (1) the perihelion distance of the particle is $> r_\mathrm{hel}$ au, then $F_>r_\mathrm{hel}=1$, and (2) the aphelion distance of the particle is $<r_\mathrm{hel}$ au, then $F_>r_\mathrm{hel}=0$.

Meteoroid dynamical models provide the position and velocity distribution of all meteoroids in the solar system. We assume that the meteoroid population is in a steady-state, i.e. the sources produce the same amount of meteoroids that are lost in sinks (impacts with planets, disintegration close to the sun, escape from the solar system on a hyperbolic orbit, loss in the collision). Knowing the diameter of each meteoroid and assuming a particular size-frequency distribution we can easily calculate $\Sigma_\mathrm{ZC}$
\begin{equation}
\Sigma_\mathrm{ZC}= \sum_{i=1}^{N_\mathrm{met}(D)}F_>r_\mathrm{hel} \frac{\pi D^2}{4}, 
\end{equation}
where $N_\mathrm{met}(D)$ is the number of meteoroids per size bin depending on the size-frequency distribution described in the following section.

\subsection{Size-frequency distribution (SFD)}
In order to put meteoroid simulations of different sizes together we apply a size-frequency distribution (SFD) to each dust population when the meteoroids are ejected. The original SFD is not conserved during the meteoroids' dynamical evolution since many effects like the Poynting-Robertson drag, radiation pressure, the mean-motion resonance capture rates or the secular resonance acting times depend on the meteoroid size and density. 

Here, we represent the meteoroid SFD of particles with diameter $D$ using a single power-law:
\begin{eqnarray}
dN &=& N_0 (\alpha -1) \left( \frac{D_\mathrm{max}}{D} \right)^{\alpha} \frac{dD}{D_\mathrm{max}},\\
N(>D) &=&  N_0 \left( \frac{D_\mathrm{max}}{D}  \right)^{\alpha-1},
\end{eqnarray}
where $N$ is the number of particles, $N_0$ is a calibration parameter, $\alpha > 0$ is the  power law index, $D_\mathrm{max}$ is the maximum diameter assumed in our model. Using any given power law we can then derive the total mass of the dust cloud:
\begin{equation}
    M_\mathrm{tot}=M_\mathrm{max}\frac{N_0(\alpha-1)}{4-\alpha}\left[ 1 - \left( \frac{D_\mathrm{min}}{D_\mathrm{max}}\right)^{4-\alpha} \right],
\end{equation}
where $M_\mathrm{max}=\pi/6 D_\mathrm{max}^3 \rho$ and $D_\mathrm{min}$ is the diameter of smallest particle in our model. Similarly we can derive the total particle cross-section:
\begin{equation}
    \Sigma_\mathrm{tot}=\Sigma_\mathrm{max}\frac{N_0(\alpha-1)}{3-\alpha}\left[ 1 - \left( \frac{D_\mathrm{min}}{D_\mathrm{max}}\right)^{3-\alpha} \right],
\end{equation}
where $\Sigma_\mathrm{max}=\pi D_\mathrm{max}^2/4$.

\bibliography{papers}

\end{document}